\documentclass[twocolumn,prl,aps,superscriptaddress]{revtex4}
\usepackage{latexsym}
\usepackage{amssymb}
\usepackage[dvips]{graphicx}

\begin{document}

\title{How free are Composite Fermions at $\nu = \frac{1}{2}$? A NMR investigation}

  \author{N.  Freytag}\email[corresponding author:~]{N.Freytag@fkf.mpg.de}
   \affiliation{Grenoble High Magnetic Field Laboratory, MPI-FKF and
    CNRS, B.P.~166, 38042 Grenoble Cedex 9, France}
   \affiliation{Max Planck Institute for Solid State
   Research, Heisenbergstr.~1, 70569 Stuttgart, Germany}

  \author{M. Horvati\'{c}} \affiliation{Grenoble High Magnetic Field
    Laboratory, MPI-FKF and CNRS, B.P.~166, 38042 Grenoble Cedex 9,
    France}

  \author{C. Berthier} \affiliation{Grenoble High Magnetic Field
    Laboratory, MPI-FKF and CNRS, B.P.~166, 38042 Grenoble Cedex 9,
    France} \affiliation{Laboratoire de Spectrom\'{e}trie Physique,
    Universit\'e J. Fourier, B.P.~87, 38402 St.~Martin
    d'H\`{e}res, France}

  \author{M. Shayegan} \affiliation{Department of Electrical
    Engineering, Princeton University, Princeton NJ~08544, USA}

  \author{L.P. L\'evy} \affiliation{Grenoble High Magnetic Field
    Laboratory, MPI-FKF and CNRS, B.P.~166, 38042 Grenoble Cedex 9,
    France} \affiliation{Institut Universitaire de France and
    Universit\'e J. Fourier, B.P.~41, 38402 St.~Martin
    d'H\`{e}res, France} \date{\today}


\begin{abstract}
NMR measurements of the electron spin polarization have been performed on a 2D electron system at and around half-filled lowest Landau level. Comparing the magnetic field and the temperature dependence of the spin polarization to models of free and interacting composite fermions (CF), we confirm the existence of a Fermi sea and show that residual interactions are important. Independent measurements of the CF effective mass, $g$-factor and Fermi energy are obtained from the thermal activation of the spin polarization in tilted fields.  The filling factor dependence of the spin polarization for $\frac{2}{5} < \nu < \frac{2}{3}$ reveals a broken particle-hole symmetry for the partially polarized CF Fermi-sea.
\end{abstract}

\maketitle




In high magnetic fields, a two-dimensional electron gas exhibits fractional quantum Hall states when the Landau level filling factor is equal to $\nu = \frac{p}{2pq \pm 1}$ or $1-\frac{p}{2pq \pm 1}$ ($p,q \, \epsilon \, \mathbb{N}$).  These highly correlated electronic states can be viewed as the integer quantum Hall states $\nu^\star = p$ of ``composite fermions'' (CF) \cite{CF} moving in an averaged reduced magnetic field $B^\star = B-2q\phi_0/{\cal A}$ ($\phi_0$ is the flux quantum, $\cal A$ is the sample area and $\frac{1}{\nu^\star} = \frac{1}{\nu}-2q$).  At half filling, $\nu = \frac{1}{2}$ ($q = 1$), CFs experience no magnetic field and form a Fermi sea.  Resonances in magneto-transport experiments \cite{MT} close to $\nu = \frac{1}{2}$ have shown that CFs are physical particles undergoing cyclotron motion in the reduced field $B^\star$: today most experimental results in the fractional quantum Hall regime can be easily visualized with this mapping onto nearly free CF particles in a reduced field.  On the other hand, quantitative comparison between experiments and theory have been more challenging: in spite of their physical relevance, residual Coulomb interactions, fluctuations and disorder are difficult to incorporate in the theory.

At low magnetic fields, CF spin degrees of freedom are also relevant since the system is not fully polarized.  For a free CF Fermi gas, the spin polarization $\cal P$ is determined by the difference in area of the spin-$\uparrow$ and spin-$\downarrow$ Fermi disks. Experimentally, $\cal P$ can be tuned by tilting the sample with respect to the magnetic field, keeping the perpendicular component to the 2D gas (and hence $\nu$) constant.  In this Letter, we probe the electron spin polarization in the vicinity of half filling using a nuclear magnetic resonance (NMR) measurement of the Knight shift of $^{71}$Ga nuclei in a GaAs/GaAlAs multiple quantum well (QW) sample  with electron density $n = 8.5 \, 10^{10}$~cm$^{-2}$ (M280), previously used in other studies \cite{1nu2paper,2nu3paper,1nupaper,thesis}.  Comparing this experimental data to a free CF model and to the Hamiltonian theory \cite{Shankar01} of CFs, which incorporates the residual Coulomb interactions, the existence of a thermodynamic Fermi sea for CF is established, with an effective mass that is \emph{independent} of the Zeeman energy.




NMR is a very sensitive {\em direct} measurement of the spin polarization $\cal P$ of electrons in QWs, since the Fermi contact interaction ${\cal H} \propto \sum_{ij} {\cal S}_i \cdot {\cal I}_j\delta({\vec r}_i-{\vec R}_j)$ between itinerant electron spins ${\cal S}_i$ at position ${\vec r}_i$ and nuclear spins ${\cal I}_j$ at ${\vec R}_j$ shifts the resonance frequency of $^{71}$Ga nuclei in the QWs by a magnetic hyperfine shift $K_{{\rm S}}$ proportional to ${\cal P}$ \cite{1nu2paper,2nu3paper,1nupaper,thesis}.
The NMR signal from the barriers, where the conduction band is empty, is used as a zero-shift reference.  The spin polarization is inferred from $K_{{\rm S}}$ as ${\cal P}(\nu ,T,\theta )=\frac{K_{{\rm S}}(\nu ,T,\theta )}{K_{{\rm S}}({\cal P}=1)}$, where $K_{{\rm S}}({\cal P}=1)$ is the maximum shift measured in this sample (\textit{e.g.}\ at low temperature for the ferromagnetic state at $\nu = \frac{1}{3}$). We used the same experimental setup as in previous studies \cite{1nu2paper,2nu3paper,1nupaper,thesis}.




\begin{figure}
\centering
\includegraphics*[width=0.9\columnwidth]{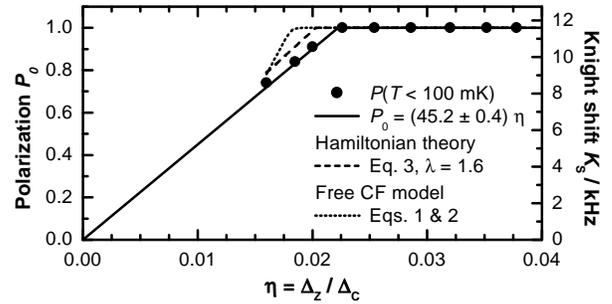}
\caption{The spin polarization (left axis) and Knight shift (right axis) in the limit of vanishing temperature versus the ratio of Zeeman and Coulomb energy $\eta = \frac{\Delta_{\rm Z}}{\Delta_{\rm C}}$ at $\nu=\frac{1}{2}$. The error bars are of the order of the symbol size.  The solid line is a linear fit to the data, while the dashed and dotted lines are the predictions of the Hamiltonian theory and the free CF model, respectively.}
\label{Fig:PvsEta}
\end{figure}

\begin{figure*}
\centering 
\includegraphics*[width=0.9\textwidth]{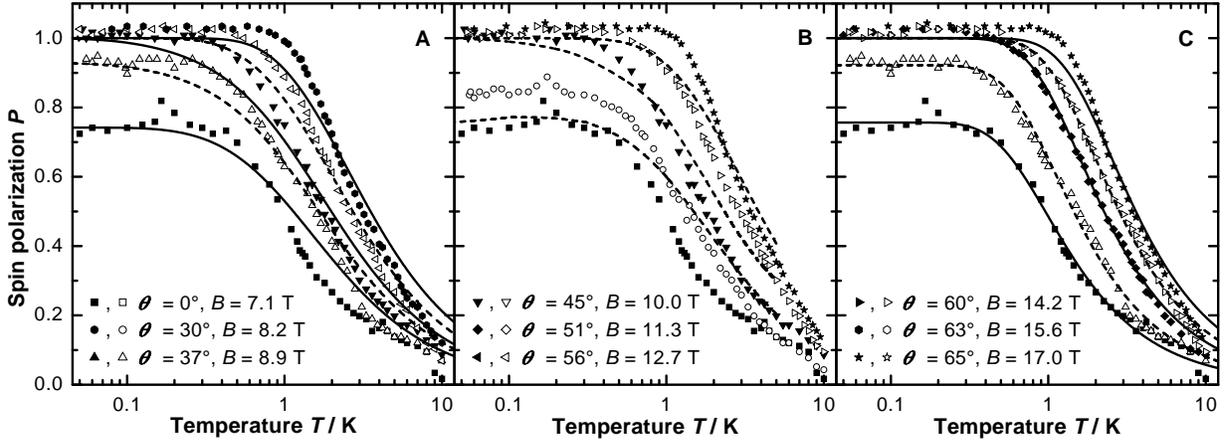} \caption{The $T$-dependence of the spin polarization at $\nu = \frac{1}{2}$ is compared to the free CF model (Eq.~\ref{Eq:PvsT1/2}) with fixed $m_{\rm P}^\star = 1.67$ (A), to the prediction of the Hamiltonian theory (B) and the best fits to a simple two level model (Eq.~\ref{Eq:tanh1/2}).  The corresponding spin-gaps $\Delta$ are shown in Fig.~\ref{Fig:gfactor}. For better visibility each panel contains a choice of data and every other symbol is open.}
\label{Fig:PvsT}
\end{figure*}

In Fig.~\ref{Fig:PvsEta}, we present the dependence of ${\cal P}_0$ on the ratio of Zeeman to Coulomb energy $\eta =\frac{\Delta_{\rm Z}}{\Delta_{\rm C}}$ at filling factor $\nu = \frac{1}{2}$
 \footnote{Here $\Delta_{\rm C} = \frac{1}{4 \pi \varepsilon_0 \varepsilon_r} \sqrt{\frac{e B}{\hbar}}$ and no correction factors due to finite width have been taken into account, while the $\Delta_{\rm Z}$ has been computed using $|g^*| = 0.44$.}. ${\cal P}_0$ is the low temperature saturation value of the spin polarization ${\cal P}(T)$ shown in Fig.~\ref{Fig:PvsT}, while $\eta$ is changed by tilting the sample, keeping $\nu$ at $\frac{1}{2}$.  The saturation value $K_{\rm S}({\cal P} = 1) = 11.6 \textrm{ kHz}$ agrees with the one previously determined \cite{2nu3paper}.  Until full polarization is reached, the data fall on a line extrapolating to ${\cal P}_0(\eta = 0) = 0$ (vanishing electron density $n$ and $\Delta_{\rm Z}$). The solid line is a linear fit, yielding a critical value of $\eta_{\rm c} = 0.022$ for a fully polarized CF Fermi-sea (FS). This smooth dependence of ${\cal P}_0$ on $\eta$ is a proof for the existence of FSs for both spin directions. In the lowest Landau level the kinetic energy is quenched and there are two highly degenerate spin levels separated by $\Delta_{\rm Z}$. At odd denominator filling factors, this degeneracy is partially reduced by the formation of CF Landau levels and only  particular, unmagnetizable states are energetically favorable, which leads to a step-like function ${\cal P}(\eta)$ \cite{Kukushkin,2nu3paper}. The appearance of a magnetizable state at $\nu = \frac{1}{2}$ is a signature of the FSs for both spin directions.

At $\eta_{\rm c}$, the Zeeman and CF Fermi energies are identical.  Thus we find, assuming the $g$-factor of CFs to be the same as for electrons ($|g^*| = 0.44$), $\varepsilon_{\rm F}^\star = \Delta_{\rm Z}(\eta_{\rm c})\approx 2.93 \textrm{ K}$. The CF Fermi energy $\varepsilon_{\rm F}^{\star} = \frac{n}{D}$ can be specified by an effective mass $m_{\rm P}^\star$, through the CF density of states $D = \frac{m^{\star}_{\rm P} m_e}{2\pi\hbar^2}$ at the Fermi level ($m_e$ is the bare electron mass and not the GaAs band mass). Using $B_\perp = \frac{2 h}{e}n$, we find $\varepsilon_{\rm F}^\star = \frac{e\hbar}{2m_{\rm P}^\star m_e}\, B_\perp \approx \frac{4.77}{m_{\rm P}^\star}$~K, and determine in this way the effective mass of CFs at $\eta_{\rm c}$ to be $m_{\rm P}^\star(\eta_{\rm c}) \approx 1.63$.

To understand the linear dependence of ${\cal P}_0$ on $\eta$ observed in Fig.~\ref{Fig:PvsEta}, we compute ${\cal P}(T,B)$ for a fixed number of CFs. Assuming a quadratic dispersion $\epsilon(k)=\frac{k^2}{4\pi D}$, the $T$-dependence of the spin polarization is \cite{1nu2paper,OPNMR}:
\begin{equation}
\label{Eq:PvsT1/2}
{\cal P}(T, B) = \frac{1}{\varepsilon_{\rm F}^{\star}}
  \left[\Delta_{\rm Z}  -  2k_{\rm B}T\,
  \mathrm{arsinh}\!\!\left(\!\frac{\sinh(\frac{\Delta_{\rm Z}}{2k_{\rm B}T})}
  {\exp (\frac{n}{2Dk_{\rm B}T})}\!\right)\! \right],
\end{equation}
which reduces to ${\cal P}_0(B) = \min \left \{ \Delta_{\rm
    Z}/\varepsilon_{\rm F}^{\star}, \, 1\right \}$.

Hence below $\eta_c$, ${\cal P}_0 = \frac{m^{\star}_{\rm P}m_e \Delta_{\rm Z}}{4\pi\hbar^2 n}$ is proportional to $\Delta_{\rm Z}$.  In the data shown in Fig.~\ref{Fig:PvsEta}, ${\cal P}_0$ varies linearly with $\eta \propto \Delta_{\rm Z}$. Thus $m^{\star}_{\rm P}$ must be constant, independent of $B$ at a given $n$ \footnote{In Ref.~\onlinecite{1nu2paper} it has been shown that $m_{\rm P}^\star$ is nevertheless proportional to $1/\sqrt{B_\perp}$.}.  Mapping the fractional gaps onto $\hbar \omega_{\rm CF} \left(n + \frac{1}{2}\right)$, Park and Jain \cite{Park98} find a field dependent effective mass $m^\star_{\rm P} \propto \eta$:  for the parameters relevant to GaAs ($|g^*| = 0.44$, $\epsilon_r = 12.7$), they determine
\begin{equation}
  \label{eq:ParkJain}
  m^{\star}_{\rm P} = 0.66 \, \frac{B/\textrm{T}}
{\sqrt{B_\perp/\mathrm{T}}}\propto \eta \, .
\end{equation}
For sample M280, \textit{i.e.}\ at $B_\perp = 7.1$~T, Eq.~\ref{eq:ParkJain} predicts $m^{\star}_{\rm P} = 1.76 / \cos (\theta)$, where $\theta$ is the sample tilt angle.  The $\theta=0$ value compares reasonably to the experimental value $m^{\star}_{\rm P} = 1.63$ determined above.  Nevertheless, the $B$-dependence of $m_{\rm P}^\star$ given by Eq.~\ref{eq:ParkJain} would lead to a nonlinear dependence of ${\cal P}_0$ on $\eta$, shown in Fig.~\ref{Fig:PvsEta} as dotted line and is not consistent with the observed behavior.

In contrast to the free CF model, the Hamiltonian theory \textit{predicts} ${\cal P}_0$ to be linear with $\eta$:  Eq.~153 in \cite{Shankar01} reads
\begin{equation}
  \label{eq:PvsEtaShankar}
  {\cal P}_0 = \frac{0.13 \, \sqrt{B_\perp}\,
\lambda^{7/4}}{\cos \theta}  \propto \eta \, ,
\end{equation}
where $B_\perp$ is expressed in Tesla and $\lambda$ is the finite thickness parameter in the Zhang-Das~Sarma potential, determined to be $\lambda = 1.6\, \ell_B$ by fitting one point of ${\cal P}(T,\theta=0)$. From Fig.~\ref{Fig:PvsEta}, we see that Eq.~\ref{eq:PvsEtaShankar}, which lies slightly above the experimental ${\cal P}(\eta)$, is qualitatively correct.

We now turn to the temperature dependence of $\cal P$. So far, $m_{\rm P}^\star$ has been determined to be constant for $\eta < \eta_{\rm c}$. Using Eq.~\ref{Eq:PvsT1/2}, we can test whether this holds also for $\eta > \eta_{\rm c}$.  Furthermore, the depolarization at finite $T$ is governed by the Zeeman energy: an independent measurement of the CF $g$-factor can be inferred.

In Fig.~\ref{Fig:PvsT}, we present the temperature dependence ${\cal P}(T,\nu = \frac{1}{2})$ for tilt angles $0^\circ \leq \theta \leq 65^\circ$, corresponding to magnetic fields between $7.1\textrm{~T}$ and $17 \textrm{~T}$.
As has already been shown in Fig.~\ref{Fig:PvsEta}, for tilt angles $\theta \geq 45^\circ$ ($\eta > 0.022$) the polarization saturates at ${\cal P} = 1$ at low $T$: the ground state is fully spin polarized. For lower tilt, however, the CF FS is only partially polarized.

\begin{figure}
\centering
\includegraphics*[width=0.9\columnwidth]{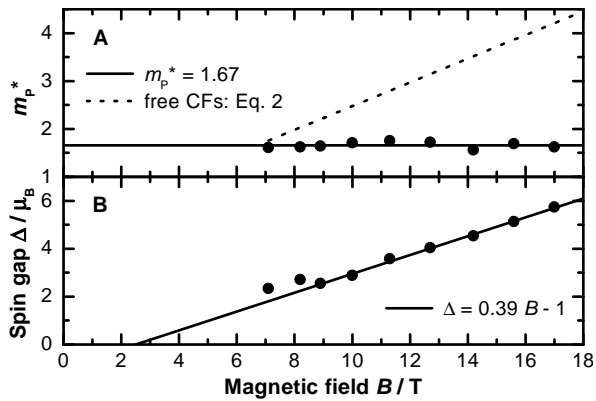}
\caption{(A) The polarization masses obtained from fits using Eq.~\ref{Eq:PvsT1/2} to our ${\cal P}(T)$ data (Fig.~\ref{Fig:PvsT}) versus magnetic field.  The solid line corresponds to $m_{\rm P}^\star = 1.67$, while the dashed line is the prediction of the free CF model (Eq.~\ref{eq:ParkJain}).  (B) The magnetic field dependence of the spin-flip gap in units of the Bohr magneton $\mu_{\rm B}$ as extracted from the fits in Fig.~\ref{Fig:PvsT}C.  The slope directly yields the CF $g$-factor.}
\label{Fig:gfactor}
\end{figure}

We now compare ${\cal P}(T,\theta)$ with the free CF model. The predictions by  Eq.~\ref{Eq:PvsT1/2}, taking a fixed mass $m_{\rm P}^\star = 1.67$ are shown in Fig.~\ref{Fig:PvsT}A.  Considering that $m_{\rm P}^\star$ is the only adjustable parameter, these curves fit quite satisfactory.  The thermal depolarization predicted by the Hamiltonian theory is shown in Fig.~\ref{Fig:PvsT}B and is roughly similar to the free CF model.  

Whereas the agreement between these models and the experiment is reasonable for all $T$ at high $\eta$, it is only fair for $T > 2$~K, when ${{\cal P}_0} < 1$ \footnote{We remark that in Ref.~\onlinecite{OPNMR} the $\nu = \frac{1}{2}$ OPNMR data has been analyzed using a fit in which Eq.~\ref{Eq:PvsT1/2} is modified by a mean field interaction term. Applied to our NMR data, this systematically degrades the agreement between fits and data.}.  This is not surprising since the CF model breaks down at high energy:  the assumption of parabolic bands leading to Eq.~\ref{eq:ParkJain} is questionable for energies exceeding the largest gap ($\Delta_{\nu=1/3}$), where CFs are expected to ``dissociate''. In fact, fits are improved significantly by limiting the CF dispersion to this finite bandwidth (using \emph{e.g.}\ $\epsilon(k) = \Delta_{1/3} \tanh^2(k\ell_{\rm B})$, not shown here). Nonetheless this does not change the qualitative arguments presented here.

In the limit of large magnetic fields, Eq.~\ref{Eq:PvsT1/2} becomes
\begin{equation}
\label{Eq:tanh1/2}
  {\cal P}(T,B \gg B_{\rm c}) \approx {\rm tanh}\left (
\frac{\Delta}{2 k_{\rm B}T}\right )\, ,
\end{equation}
where the spin-gap $\Delta \approx g^{\star} \mu_{\rm B}B - \varepsilon_{\rm F}^\star$ is independent of $m_{\rm P}^\star$ (since $m_{\rm P }^\star(B) \approx 1.67$, $\varepsilon_{\rm F}^\star \propto \frac{1}{m_{\rm P }^\star}$ is an irrelevant constant).  In Fig.~\ref{Fig:PvsT}C, we analyze the data using this fit, even for $B<B_{\rm c}$ where this approximation does \textit{not} hold. In the latter case the ground state polarization is partial ($\theta \leq 37^\circ$) and we add an additional fitting parameter:  ${\cal P}_0$. 
Eq.~\ref{Eq:tanh1/2} reproduces the data for \textit{all} temperatures with high accuracy.  The corresponding spin-flip gaps $\Delta$ are shown in Fig.~\ref{Fig:gfactor}B.
For large magnetic fields ($\Delta_{\rm Z} \gg \varepsilon_{\rm F}^\star$) this can be understood, since the CF spectrum ``looks like'' a two-level system. Nevertheless, for low $B$ the good agreement achieved by this fit is surprising. It is far better than those obtained with the free CF model or the Hamiltonian theory (Figs.~\ref{Fig:PvsT}A \& B). This behavior remains to be understood.

Since ${\cal P}(T)$ depends mostly on the Zeeman energy, it can be used to determine the $g$-factor of CFs ($g^\star$).  In contrast to transport measurements \cite{Stoermer}, $\Delta_{\rm Z}$ is here determined directly at $\nu = \frac{1}{2}$ and gives $g^\star$ independently from the effective mass $m_{\rm P}^\star$. Since in Fig.~\ref{Fig:gfactor}B the spin-gap is plotted in units of the Bohr magneton, the slope of $\Delta(B)$ directly yields $g^\star$.

A linear fit to $\Delta(B)$ for $B>B_{\rm c}$ yields $g^\star = 0.39$, very close to the value for bulk GaAs $|g^*|=0.44$.  We performed a self-consistent calculation of the band structure, leading to an effective $g$-factor for electrons $|g^*| \approx 0.42 - 0.002 \, B/\textrm{T}$, almost identical to the bulk GaAs value.  Hence, the CF $g$-factor is essentially the \textit{same} as for electrons in GaAs, justifying \textit{a posteriori} the determination of $\varepsilon_{\rm F}^\star$ and $m_{\rm P}^\star$ using $g^\star = 0.44$.
Here we remark that a strong enhancement of $g^\star$ has been reported ($g^\star \approx 0.6$ at $\nu = \frac{3}{2}, \frac{1}{4}$ \cite{Stoermer} and $g^\star\approx 1.1$ at $\frac{1}{2}$ \cite{Kukushkin}). 
These determinations of $g^\star$ are in contradiction with this NMR study, which is mostly a thermodynamic measurement of $\cal P$.
We cannot exclude that the strong discrepancy with Ref.~\onlinecite{Kukushkin} could result from dynamically polarized nuclei leading to an increased Zeeman energy through the hyperfine field.

\begin{figure}[t]
\centering 
\includegraphics*[width=0.9\columnwidth]{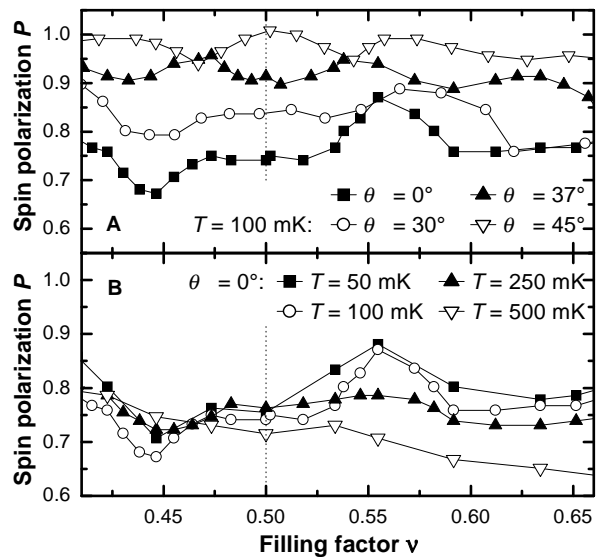}
\caption{The filling factor dependence of the polarization for $\frac{2}{5} < \nu < \frac{2}{3}$ at $T = 100$ mK and various tilt angles $\theta$ (A) and at $\theta = 0^{\circ}$ and temperatures between $50$ and $500$ mK (B).  The thin lines are to guide the eye.}
\label{Fig:PvsNu}
\end{figure}

We now consider the filling factor dependence of $\cal P$ around $\nu = \frac{1}{2}$ shown in Fig.~\ref{Fig:PvsNu}. Here, we are interested in small deviations from $\nu = \frac{1}{2}$ in order to investigate the breakdown of the FSs at half filling due to the quantization of the kinetic energy of CFs.  In this range of magnetic fields around $\nu = \frac{1}{2}$, the longitudinal resistivity is relatively flat.  Fig.~\ref{Fig:PvsNu}A depicts ${\cal P}(\nu)$ for $\frac{2}{5} < \nu < \frac{2}{3}$ for different $\theta$ at $T = 100$~mK,
while Fig.~\ref{Fig:PvsNu}B presents data at temperatures from $50$~mK to $500$~mK at $\theta = 0^\circ$.  Focusing on the $\theta = 45^{\circ}$ data (highest field), we observe nearly full polarization over the whole range of $\eta$, with a slight depolarization at $\nu = \frac{1}{2} \pm 0.04$.  These dips in $\cal P$ are symmetric with respect to $\nu=\frac{1}{2}$ as expected due to particle-hole symmetry for large Zeeman energy \footnote{Charge conjugation maps $\nu \to 1-\nu$ for large $\Delta_{\rm Z}$, while $\nu \to 2-\nu$ for small $\Delta_{\rm Z}$.}. The situation is clearly different at lower tilt angles:  for the untilted sample, we observe features at $\nu = \frac{1}{2} \pm 0.06$ with decreased ${\cal P}$ at \emph{higher} and increased ${\cal P}$ at \emph{lower} Zeeman energy.  For larger tilt angles, these asymmetric features smear out and disappear completely at $\theta = 45^{\circ}$ when full polarization is reached.

In Fig.~\ref{Fig:PvsNu}B, it becomes clear that these features only develop at very low $T$.  While at 500~mK, we observe a monotonic increase of $\cal P$ for $\frac{2}{5} < \nu < \frac{2}{3}$, the features start to appear at $T = 250$~mK and are most pronounced at lowest $T$:  \emph{e.g.}\ at $\nu = 0.44$, the $\cal P$ actually drops when the temperature is decreased.  We conclude that our data strongly suggest a violation of the particle-hole symmetry around $\nu=\frac{1}{2}$ for a partially polarized FS.

The hyperfine coupling between electron and nuclear spins is not only responsible for $K_{\rm S}$ but also for the nuclear spin lattice relaxation time $T_1$.

\begin{figure}
\centering 
\includegraphics*[width=0.9\columnwidth]{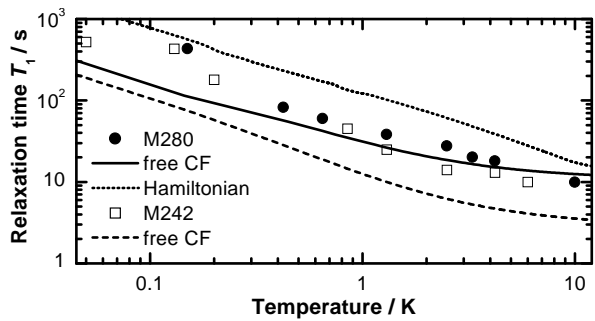}
\caption{The nuclear spin lattice relaxation time $T_1$ versus temperature for the untilted samples M280 and M242 at $\nu=\frac{1}{2}$.  The solid and dashed lines are $T_1(T)$ in the free CF (Eq.~\ref{eq:T1}), and the dotted line in the Hamiltonian theory (only M280).}
\label{Fig:T1vsT}
\end{figure}

We compute $T_1$ using the free CF model and find for $^{71}$Ga nuclei in the center of the quantum wells:
\begin{equation}
  \label{eq:T1}
T_1^\star = \frac{1 + \frac{\exp\left(\frac{\Delta_Z}{2k_{\rm B}T}\right)}
{\sqrt{\exp\left(\frac{\varepsilon_{\rm F}}{k_{\rm B} T}\right) +
\sinh^2\left(\frac{\Delta_Z}{2k_{\rm B}T}\right)} -
\cosh\left(\frac{\Delta_Z}{2k_{\rm B}T}\right)}} {\frac{4\pi}{\hbar^3} \left
(\frac{ m_{\rm P}^\star K_{\rm S}({\cal P}=1)}{n}\right)^2 k_{\rm B}T} \, .
\end{equation}

For sample M280 as well as a higher density sample M242 \cite{1nu2paper,2nu3paper,1nupaper,thesis}, we have measured $T_1(\nu = \frac{1}{2}, \theta = 0^\circ)$ (shown in Fig.~\ref{Fig:T1vsT}) using a saturation-recovery method \cite{1nupaper}.  The solid (dashed) line is the prediction of the free CF model, Eq.~\ref{eq:T1}, for M280 (M242), roughly three times shorter than the experimental $T_1$.  Also included is the calculation for M280 by Shankar using the Hamiltonian theory \cite{Shankar01} (dotted line), being twice as long as $T_1(\textrm{M280})$.  
Note that the discrepancy for free CFs is even larger when other relaxation processes are included, while Shankar's calculation might become of the right order if relaxation due to edge-states and impurities were included.




In conclusion, the experiments at $\nu = \frac{1}{2}$ presented in this Letter prove the existence of CF Fermi seas for both spin directions, shifted by $\Delta_{\rm Z}$. The free CF model provides a reasonable description, provided that the effective mass $m^\star_{\rm P}$ is constant with $\Delta_{\rm Z}$, a feature which emerges naturally from the Hamiltonian theory. From the temperature dependence of $\cal P$, the CF-Fermi energy and $g$-factor have been determined, while the $T_1(T)$ measurement reveals the importance of residual interactions. The $\nu$-dependence around half filling suggests a violation of particle-hole symmetry for partially polarized FS.




We gratefully acknowledge R. Shankar for many fruitful discussions and for providing the theoretical curves shown in Figs.~\ref{Fig:PvsT}B and \ref{Fig:T1vsT}, as well as J.H. Smet and K.\ v.\ Klitzing for carefully reading the manuscript. This work was partially supported by the NSF.



\end{document}